\documentclass[conference]{IEEEtran}
\IEEEoverridecommandlockouts
\usepackage{cite}
\usepackage{amsmath,amssymb,amsfonts}
\usepackage{algorithmic}
\usepackage{graphicx}
\usepackage{textcomp}
\usepackage{orcidlink}

\def\BibTeX{{\rm B\kern-.05em{\sc i\kern-.025em b}\kern-.08em
    T\kern-.1667em\lower.7ex\hbox{E}\kern-.125emX}}

\renewcommand\IEEEkeywordsname{\\ Keywords}   

                  \usepackage[english]{babel}
                  \usepackage[utf8]{inputenc}
                  \usepackage{fancyhdr}
                  \fancypagestyle{firstpage}{
                  }

\begin{document}

\title{Streamlining Brain Tumor Classification with Custom Transfer Learning in MRI Images
}


\author{\IEEEauthorblockN{\textsuperscript{1}Javed Hossain,  \textsuperscript{2}Md. Touhidul Islam, {\textsuperscript{3} Md. Taufiqul Haque Khan Tusar}}
\IEEEauthorblockA{\textit{\textsuperscript{1,3}Dept. of Computer Science and Engineering, City University, Dhaka 1216, Bangladesh}}
\IEEEauthorblockA{\textit{\textsuperscript{2}Dept. of Computer Science and Engineering, East West University, Dhaka 1212, Bangladesh}\\
\textsuperscript{1}javed.cucse@gmail.com, \textsuperscript{2}touhid.cse@cityuniversity.edu.bd, \textsuperscript{3}taufiqkhantusar@gmail.com\\
\textsuperscript{1}0000-0003-3303-7371 \orcidlink{0000-0003-3303-7371}, \textsuperscript{2} 0000-0003-2377-5071 \orcidlink{0000-0003-2377-5071}, \textsuperscript{3}0000-0002-5586-6819 \orcidlink{0000-0002-5586-6819}}
}



\maketitle
\thispagestyle{firstpage}  

\begin{abstract}
Brain tumors are increasingly prevalent, characterized by the uncontrolled spread of aberrant tissues in the brain, with almost 700,000 new cases diagnosed globally each year. Magnetic Resonance Imaging (MRI) is commonly used for the diagnosis of brain tumors and accurate classification is a critical clinical procedure. In this study, we propose an efficient solution for classifying brain tumors from MRI images using custom transfer learning networks. While several researchers have employed various pre-trained architectures such as RESNET-50, ALEXNET, VGG-16, and VGG-19, these methods often suffer from high computational complexity. To address this issue, we present a custom and lightweight model using a Convolutional Neural Network-based pre-trained architecture with reduced complexity. Specifically, we employ the VGG-19 architecture with additional hidden layers, which reduces the complexity of the base architecture but improves computational efficiency. The objective is to achieve high classification accuracy using a novel approach. Finally, the result demonstrates a classification accuracy of 96.42\%.
\end{abstract}


\section{Introduction}
The incidence of brain tumor disease is on the rise, and it is primarily caused by the alteration of genes in the chromosomes of brain cells, leading to a change in their DNA and the formation of a new structure \cite{Rehman}. Over 120 different types of brain tumors have been identified by researchers, which can be categorized into benign (non-cancerous) and malignant (cancerous) \cite{Chelghoum}. Meningioma, pituitary adenoma, Craniopharyngioma, and Schwannoma are a few of the frequently occurring brain tumors. The occurrence of brain tumors is more prevalent in adults aged between 45 to 65 years old, and certain genetic conditions, such as tuberous sclerosis, neurofibromatosis type 1 (NS 1), neurofibromatosis type 2 (NS 2), and Turner syndrome, can increase the likelihood of developing a brain tumor \cite{Ahuja}. In the United States, it is anticipated that a primary malignant brain tumor will be detected in 25,050 adults (14,170 male and 10,880 female) this year \cite{Cancer}.

Brain tumors are typically diagnosed through MRI scans, which utilize powerful magnetic fields and radio waves to generate images of the inside of the human body. However, the task of determining the malignancy of a brain tumor from an MRI image presents a significant challenge for clinicians. The process involves a comprehensive evaluation of the image, which requires both technical expertise and a thorough understanding of the latest diagnostic techniques. This complexity is compounded by the need to accurately interpret the results of the MRI scan, given the amount of information contained within the image and the difficulty in recognizing the signs of cancer. Clinicians must have substantial experience in interpreting MRI scans, as well as a strong knowledge of diagnostic techniques, to arrive at an accurate diagnosis.  In short, the determination of the cancerous nature of a brain tumor from an MRI image is a complex and often laborious task for healthcare professionals, and this is where the integration of AI comes into play by presenting a transformative change in the working procedure of medical diagnoses, offering the potential to automate the procedure by alleviating the workload of healthcare professionals and improving patient outcomes \cite{taufiq_leukemia}.

In this paper, we developed a Convolutional Neural Network (CNN) architecture named Visual Geometry Group (VGG19) with additional dense and dropout layers by utilizing 3303 MRI image data to classify brain tumors, making it easier for clinicians in taking decision and patients to understand the diagnosis. VGG19 is a pre-trained model that comprises 19 layers in total, including 16 convolutional layers and three fully connected layers. The architecture employs 2D Convolutional layers, followed by max-pooling layers after every three to four convolutional layers \cite{Simonyan}. The three fully connected layers, including the output layer, are also utilized. Previous studies have employed machine learning algorithms, such as Support Vector Machines (SVM), K-Nearest Neighbors (KNN), and Decision Tree (DT), in conjunction with VGG19 architecture and a machine learning algorithm. However, we have decided to employ only the deep learning model and have added three extra dropout and three dense layers to the output of the VGG19 architecture. The dense layer connects every neuron to each other, allowing the transfer of output from one layer to the next as input. The dropout layer serves to eliminate unnecessary neurons, reducing both training time and computational complexity.

The current state of the field in regards to utilizing deep learning techniques for the detection of brain tumors from MRI scans is explored in several seminal papers. In a study presented in \cite{Rehman}, the authors implement a two-step model development process, which involves the use of a novel CNN architecture-based brain tumor extraction, followed by pre-trained VGG19 deep feature extraction. The authors employ a 3D CNN model for tumor segmentation and a pre-trained VGG19 model for feature extraction and selection, resulting in the highest accuracy of 98.32\%, 96.97\%, and 92.67\%. In \cite{Chelghoum}, the authors conduct an experiment using a machine learning algorithm, specifically the Naive Bayes algorithm, which is based on conditional probability, resulting in an accuracy of 91.67\%. Another paper \cite{Ahuja} presents research utilizing CNN-based pre-trained architectures, exploring the use of nine different pre-trained architectures, including AlexNet, GoogleNet, VGG16, VGG19, ResNet18, ResNet50, ResNet101, and ResNet-Inception-v2, resulting in the peak accuracy of 98.71\%. In the study \cite{Swati}, the authors employ the VGG19 architecture in their experiment and adopt a layer-wise fine-tuning strategy, adding one layer at a time and then fine-tuning the entire nineteen layers. This was done in consideration of five-fold cross-validation, with a focus on the fine-tuning technique, resulting in the highest accuracy of 96.13\%. 
Furthermore, they devised a tela-diagnostic software, facilitating real-time remote diagnosis. An alternative approach was presented in \cite{Islam}, where the authors utilized a machine learning algorithm, using Principal Component Analysis (PCA) for feature extraction and K-means clustering for tumor classification, achieving a lower execution time. However, this methodology of clustering is susceptible to becoming ensnared in local optima due to its heuristic nature. To mitigate this issue, the authors in \cite{touhid_sir_pca} integrated K-means with the Hybrid Genetic Algorithm (HGA) for the purpose of data clustering, resulting in a prediction accuracy of 94.06\%. Finally, \cite{Iqbal} presents research that utilized a combination of CNN and long short-term memory (LSTM) segmented architecture, consisting of three convolution layers, a max-pooling layer, and an LSTM layer, which served as a dense layer preceding the output layer. The highest accuracy obtained was 82\%.

\section{Methodology} \label{method}

\subsection{Data Collection} 
\begin{figure}[!hbt]
    \centering
    \includegraphics[width=0.40\textwidth]{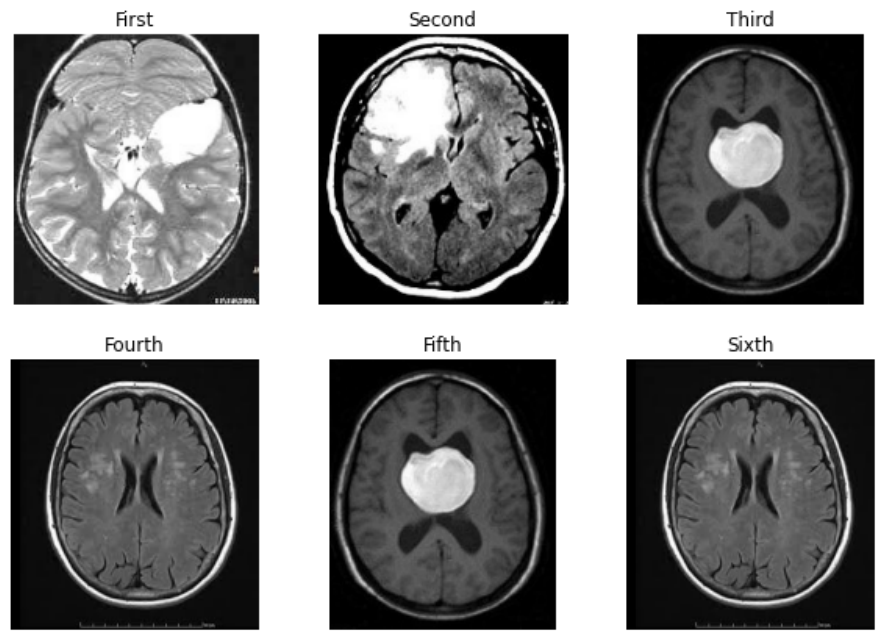}
    \caption{Samples of the MRI images on this study}
    \label{fig_data}
\end{figure}

A comprehensive collection of 3303 brain MRI images has been amassed, with 3253 images obtained from two online databases \cite{dataset_1} \cite{dataset_2} and 50 MRI images acquired through real-time scans conducted in two different hospitals located in Bangladesh. The online database comprises 1655 MRI images of individuals diagnosed with a brain tumor and 1598 images of healthy subjects, while the real-time scans encompass 30 images of tumor-affected patients and 20 images of healthy individuals. Furthermore, we have implemented image augmentation and generated 29727 images, where tumor affected and healthy subjects are 15165 and 14562 respectively. Table \ref{tab: data_distribution} shows the data distribution of collected and augmented images of our study. To facilitate subsequent procedures, the amalgamated datasets were divided into training, validation, and testing sets with a possible equal distribution of both affected and non-affected tumor samples, thus minimizing any potential biases towards a specific class. To ensure compatibility with the transfer deep learning network, the original images, which were initially collected at a resolution of $512\times512$ pixels, were resized to a reduced size of $224\times224$ pixels. Illustrative MRI images of brain tumors are shown in Fig.~\ref{fig_data}.

\begin{table}[!hbt]
\caption{Distribution of the images in the collected and augmented data}
\label{tab: data_distribution}
\begin{tabular}{cccc}
\hline
\textbf{Dataset} & \textbf{Tumor Affected} & \textbf{Healthy Subject} & \textbf{Total Image} \\ \hline
Br35H     & 1500  & 1500  & 3000  \\
BMI-BTD   & 155   & 98    & 253   \\
Realtime  & 30    & 20    & 50    \\
Augmented & 15165 & 14562 & 29727 \\ \hline
\end{tabular}
\end{table}

\subsection{Preprocessing}
\begin{figure}[!hbt]
    \centering
    \includegraphics[width=0.40\textwidth]{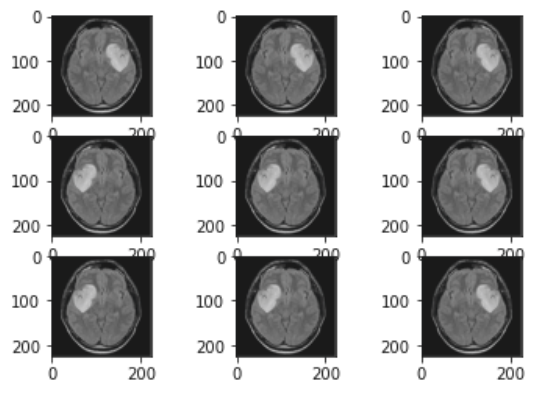}
    \caption{A sample of the augmented image}
    \label{fig_augmented}
\end{figure}

The implementation of image processing constitutes a pivotal factor in tackling the issue of image classification, and data augmentation stand out as one of its integral components. This technique encompasses the creation of a multitude of augmented images by subjecting them to transformations such as horizontal, vertical, and angular shifts. Such an approach effectively curbs the problem of overfitting, which ultimately leads to improved performance when the model is subjected to previously unseen or test data. The application of data augmentation led to the generation of a sizable corpus of 29,727 images, which comprised 15,165 images of individuals suffering from brain tumors and 14,562 images of healthy subjects. This augmentation process involved the creation of nine additional images for each original input image, thereby augmenting the model's compatibility and robustness. A sample of the augmented image is presented in Fig. \ref{fig_augmented}. Furthermore, all images were resized to a specific dimension of $224\times224$ pixels in terms of width and height. Additionally, each image underwent conversion into a numerical representation, as deep learning models usually operate on numerical values that correspond to the individual pixels in an image.

\subsection{Data Partitioning}
\begin{figure}[!hbt]
    \centering
    \includegraphics[width=0.40\textwidth]{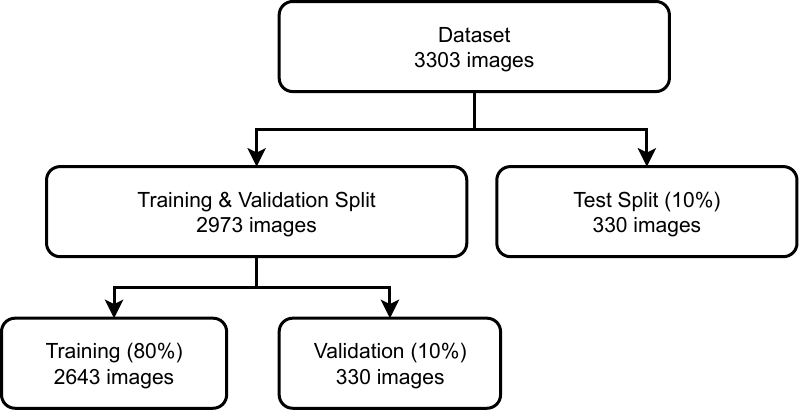}
    \caption{Data splitting approach}
    \label{fig_data_partitioning}
\end{figure}

The splitting of data holds paramount importance in the field of machine learning or deep learning, with regard to the evaluation and fortification of a model's performance. The objective of employing machine learning algorithms to construct a model is to understand pattern from real-world data that has never been encountered before and to determine a consistent methodology for its prediction or categorization. The occurrence of data leakage renders a model inadequate in its ability to perform effectively in the face of novel data in the actual world, thereby emphasizing the indispensable nature of implementing precautionary measures to preclude data leakage during the development of a model \cite{taufiq_ckd}. In our method, the data has been split into three separate groups called Training, Validation, and Testing. During the training phase, the model extracts features from the training images and develops an understanding of the same through learning. In our implementation, 80\% of the images were utilized for training, while 20\% were designated for validation and testing purposes, with an equal distribution of 10\% each for validation and testing, which has been shown in Fig. ~\ref{fig_data_partitioning}. Upon completion of the training phase, we rigorously evaluated the efficacy of our model using the test images. 

\subsection{Transfer Learning and VGG-19 Model}
In this study, we employed a CNN-based VGG-19 architecture, which was further customized with additional dropout and dense layers to enhance its performance. The integration of the dropout layer played a crucial role as it prevented the neurons in a layer from synchronizing their weight optimization, thus effectively addressing overfitting. Transfer Learning is a machine learning method where a pre-trained model that has been utilized for a specific task is utilized as the starting point for a new model that is tasked with a different problem. A domain $D = (X, P(X))$ is defined by two components, the feature space $X$ and the marginal probability distribution $P(X)$, where $X$ consists of n instances ${x_1, x_2, ....., x_n}$. The difference between the two domains is established based on either the variations in their feature spaces $(X_t \ne X_s)$ or the variations in their marginal distributions $(P(X_t) \ne P(X_s)$.

\subsection{Proposed Model}
\begin{figure}[!hbt]
    \centering
    \includegraphics[width=0.40\textwidth]{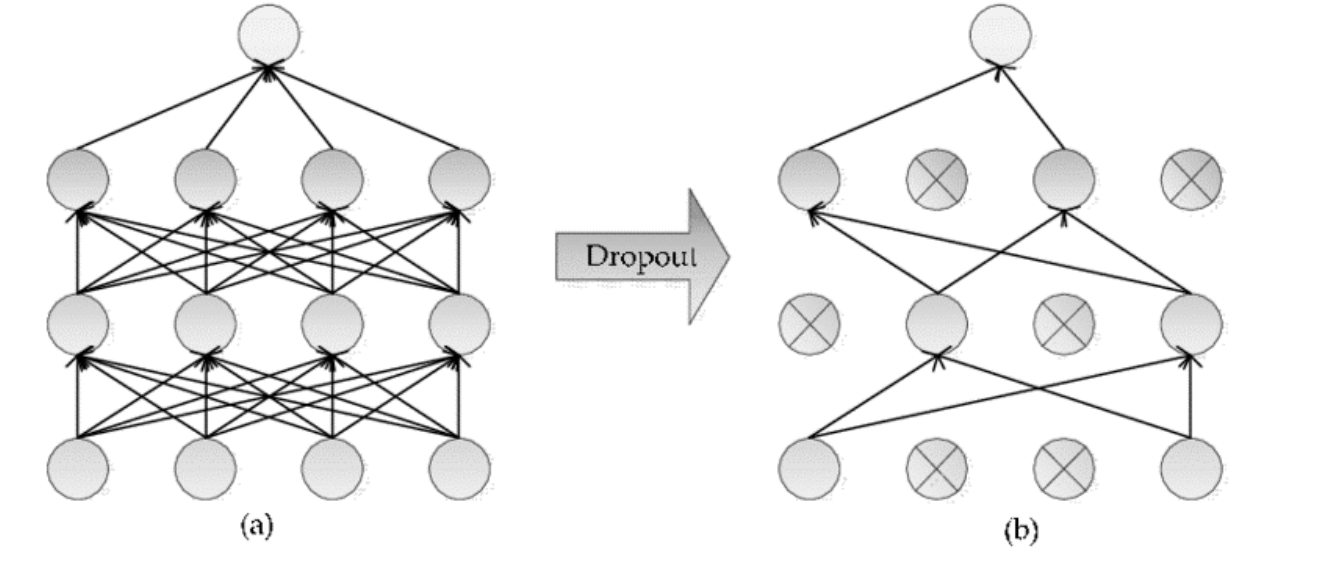}
    \caption{Regularization process of the dropout layers}
    \label{fig_dropout}
\end{figure}

We employed additional Dropout and Dense layers in our model as shown in Table \ref{tab: model_architecture}. The Dropout layer was implemented with the purpose of randomly ignoring a designated number of outputs during training, in order to decrease the mathematical computational complexity and prevent overfitting through the utilization of a regularization technique. Fig. ~\ref{fig_dropout} shows how dropout layers regularize the connections of nodes. 
\begin{table}[!hbt]
\centering
\caption{The architecture of the proposed model}
\label{tab: model_architecture}
\resizebox{\columnwidth}{!}{%
\begin{tabular}{llll}
\hline
\multicolumn{1}{|l}{\textbf{Layer}} & \multicolumn{1}{l|}{\textbf{Shape}} & \textbf{Layer} & \multicolumn{1}{l|}{\textbf{Shape}} \\ \hline
\multicolumn{1}{|l}{Input}      & \multicolumn{1}{l|}{$224\times224\times3$}   & Relu4\_2                                          & \multicolumn{1}{l|}{$28\times28\times512$} \\
\multicolumn{1}{|l}{Conv1\_1}   & \multicolumn{1}{l|}{$224\times224\times64$}  & Conv4\_3                                          & \multicolumn{1}{l|}{$28\times28\times512$} \\
\multicolumn{1}{|l}{Relu1\_1}   & \multicolumn{1}{l|}{$224\times224\times64$}  & Relu4\_3                                          & \multicolumn{1}{l|}{$28\times28\times512$} \\
\multicolumn{1}{|l}{Conv1\_2}   & \multicolumn{1}{l|}{$224\times224\times64$}  & Conv4\_4                                          & \multicolumn{1}{l|}{$28\times28\times512$} \\
\multicolumn{1}{|l}{Relu1\_2}   & \multicolumn{1}{l|}{$224\times224\times64$}  & Relu4\_4                                          & \multicolumn{1}{l|}{$28\times28\times512$} \\
\multicolumn{1}{|l}{Maxpool\_1}     & \multicolumn{1}{l|}{$112\times112\times64$}     & Maxpool4       & \multicolumn{1}{l|}{$14\times14\times512$}      \\
\multicolumn{1}{|l}{Conv2\_1}   & \multicolumn{1}{l|}{$112\times112\times128$} & Conv5\_1                                          & \multicolumn{1}{l|}{$14\times14\times512$} \\
\multicolumn{1}{|l}{Relu2\_1}   & \multicolumn{1}{l|}{$112\times112\times128$} & Relu5\_1                                          & \multicolumn{1}{l|}{$14\times14\times512$} \\
\multicolumn{1}{|l}{Conv2\_2}   & \multicolumn{1}{l|}{$112\times112\times128$} & Conv5\_2                                          & \multicolumn{1}{l|}{$14\times14\times512$} \\
\multicolumn{1}{|l}{Relu2\_2}   & \multicolumn{1}{l|}{$112\times112\times128$} & Relu5\_2                                          & \multicolumn{1}{l|}{$14\times14\times512$} \\
\multicolumn{1}{|l}{Maxpool\_2} & \multicolumn{1}{l|}{$56\times56\times128$}   & Conv5\_3                                          & \multicolumn{1}{l|}{$14\times14\times512$} \\
\multicolumn{1}{|l}{Conv3\_1}   & \multicolumn{1}{l|}{$56\times56\times256$}   & Relu5\_3                                          & \multicolumn{1}{l|}{$14\times14\times512$} \\
\multicolumn{1}{|l}{Relu3\_1}   & \multicolumn{1}{l|}{$56\times56\times256$}   & Conv5\_4                                          & \multicolumn{1}{l|}{$14\times14\times512$}\\
\multicolumn{1}{|l}{Conv3\_2}   & \multicolumn{1}{l|}{$56\times56\times256$}   & Relu5\_4                                          & \multicolumn{1}{l|}{$14\times14\times512$} \\
\multicolumn{1}{|l}{Relu3\_2}   & \multicolumn{1}{l|}{$56\times56\times256$}   & Maxpool5                                          & \multicolumn{1}{l|}{$7\times7\times512$}   \\
\multicolumn{1}{|l}{Conv3\_3}   & \multicolumn{1}{l|}{$56\times56\times256$}   & Dense6                                            & \multicolumn{1}{l|}{$1\times1\times4096$}  \\
\multicolumn{1}{|l}{Relu3\_3}   & \multicolumn{1}{l|}{$56\times56\times256$}   & Relu6                                             & \multicolumn{1}{l|}{$1\times1\times4096$}  \\
\multicolumn{1}{|l}{Conv3\_4}   & \multicolumn{1}{l|}{$56\times56\times256$}   & Drop6                                             & \multicolumn{1}{l|}{$1\times1\times4096$} \\
\multicolumn{1}{|l}{Relu3\_4}   & \multicolumn{1}{l|}{$56\times56\times256$}   & Dense7                                            & \multicolumn{1}{l|}{$1\times1\times4096$}  \\
\multicolumn{1}{|l}{Maxpool3}   & \multicolumn{1}{l|}{$56\times56\times256$}   & Relu7                                             & \multicolumn{1}{l|}{$1\times1\times4096$} \\
\multicolumn{1}{|l}{Conv4\_1}   & \multicolumn{1}{l|}{$28\times28\times512$}   & Drop7                                             & \multicolumn{1}{l|}{$1\times1\times4096$}  \\
\multicolumn{1}{|l}{Relu4\_1}   & \multicolumn{1}{l|}{$28\times28\times512$}   & Dense8                                            & \multicolumn{1}{l|}{$1\times1000$}    \\
\multicolumn{1}{|l}{Conv4\_2}   & \multicolumn{1}{l|}{$28\times28\times512$}   & Output                                            & \multicolumn{1}{l|}{$1\times1000$}    \\ \hline
                                & \multicolumn{2}{c}{\textbf{+}}                                                       &                                \\ \cline{2-3}
\multicolumn{1}{l|}{}           & \textbf{Layer}                   & \multicolumn{1}{l|}{\textbf{Shape}}               &                                \\ \cline{2-3}
\multicolumn{1}{l|}{}           & Dropout\_9 (10\% neuron)         & \multicolumn{1}{l|}{$1\times1\times1000$}                     &                                \\
\multicolumn{1}{l|}{}           & Dense\_10                        & \multicolumn{1}{l|}{$1\times1000$}                       &                                \\
\multicolumn{1}{l|}{}           & Dropout\_11(10\% neuron)         & \multicolumn{1}{l|}{$1\times1\times1000$}                     &                                \\
\multicolumn{1}{l|}{}           & Dense\_12                        & \multicolumn{1}{l|}{$1\times1\times1000$}                     &                                \\
\multicolumn{1}{l|}{}           & Dropout\_13(10\% neuron)         & \multicolumn{1}{l|}{$1\times1\times1000$}                     &                                \\
\multicolumn{1}{l|}{}           & Dense\_14 (Prediction)            & \multicolumn{1}{l|}{1 (Affected or Not Affected)} &                                \\ \cline{2-3}
\end{tabular}%
}
\end{table}

\begin{figure}[!hbt]
    \centering
    \includegraphics[width=0.40\textwidth]{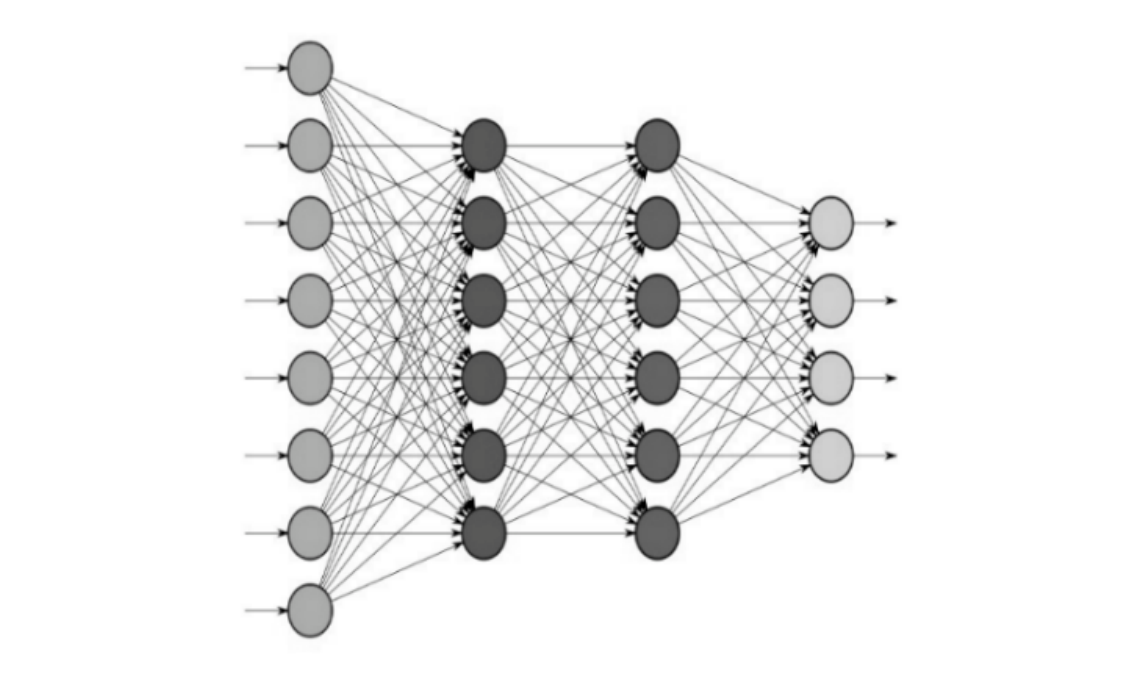}
    \caption{Architecture of the dense layers}
    \label{fig_dense}
\end{figure}

On the other hand, the Dense layer, referred to as a fully-connected layer, was integrated into our model as it enables the connection of every neuron in the previous layer to every neuron in the subsequent layer through the performance of matrix multiplication. To illustrate this, we can consider the entries in the row of a matrix to be represented as $r1, r2, ..., rn$ and the entries in the column as $c1, c2, ..., cn$. The product of the row and the column then results in a $1\times1$ matrix represented as $[r1 * c1 + r2 * c2 + ... + rn * cn]$. Fig. ~\ref{fig_dense} shows the architecture of dense layers.

\section{Results \& Discussion}
The present study was conducted on a system that uses an Intel(R) Core (TM) i5-10300H CPU operating at 2.50 GHz and featuring 8GB of RAM. The Jupyter notebook environment, which was equipped with the Python kernel and TensorFlow, was used as the computational platform for the present experiment.

\begin{figure}[!hbt]
    \centering
    \includegraphics[width=8.5cm,height=6cm]{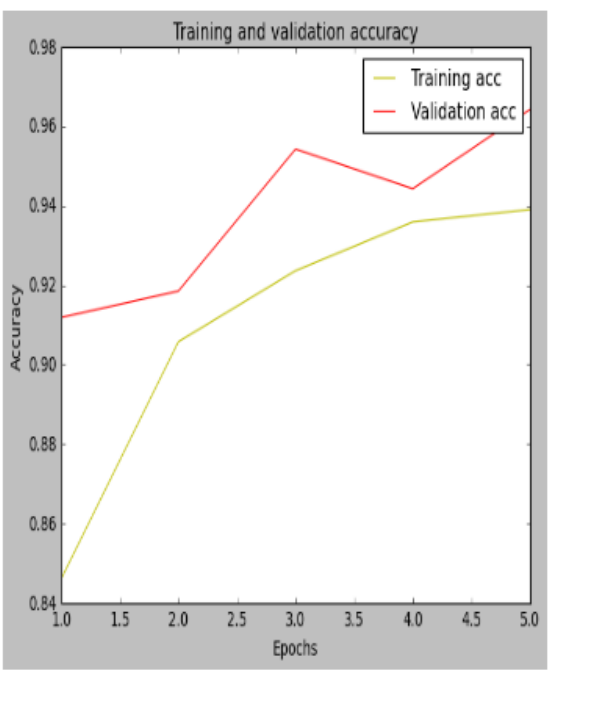}
    \caption{Training and Validation Accuracy}
    \label{fig_train_val_acc}
\end{figure}

\begin{figure}[!hbt]
    \centering
    \includegraphics[width=7.6cm,height=6cm]{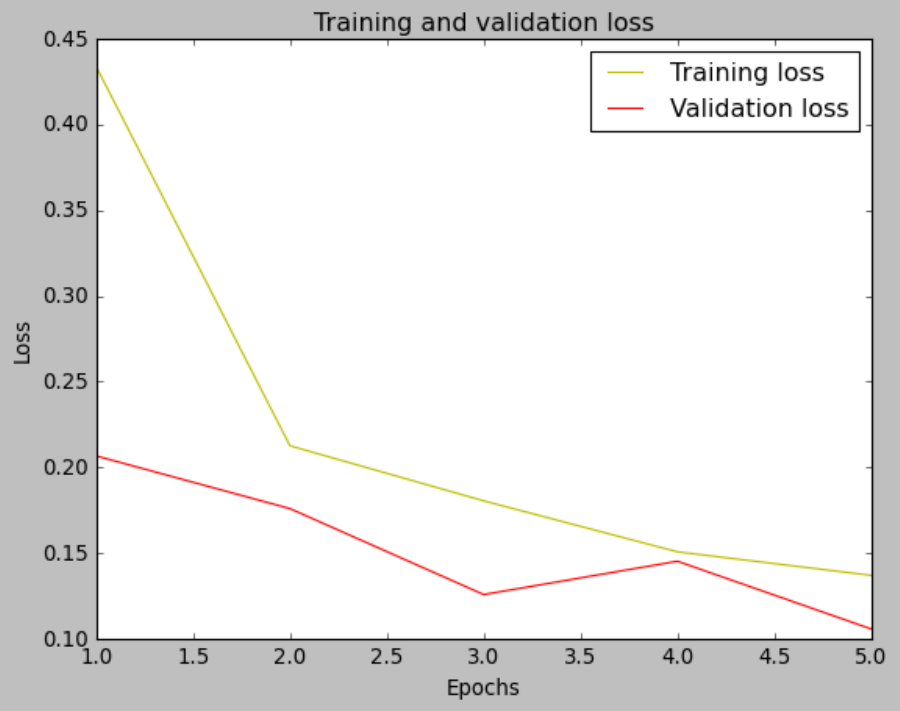}
    \caption{Training and Validation Loss}
    \label{fig_train_val_loss}
\end{figure}

Initially, the VGG-19 model was employed, and the results indicated a remarkable improvement in performance up to 10 epochs, as evidenced by the remarkable training accuracy of 99.42\% and validation accuracy of 96.25\%. Subsequently, the implementation of the proposed model resulted in a substantial increase in the training accuracy to 99.62\% and a marginal improvement in the validation accuracy to 96.42\%, as depicted in Fig. ~\ref{fig_train_val_acc} and ~\ref{fig_train_val_loss}, which represent the accuracy and loss profiles of the training and validation steps, respectively.

\begin{figure}[!hbt]
    \centering
    \includegraphics[width=0.40\textwidth]{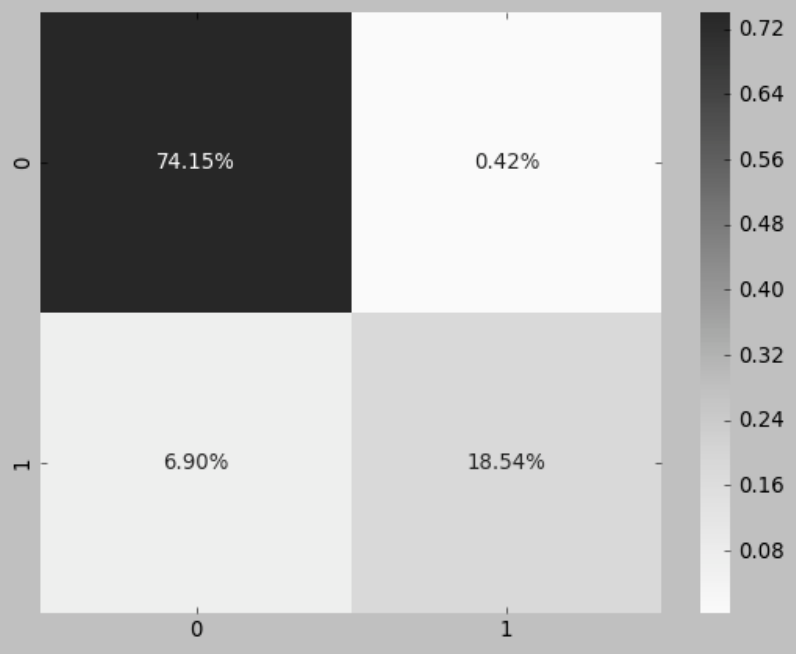}
    \caption{Confusion matrix of the test result}
    \label{fig_confusion_matrix}
\end{figure}

\begin{figure}[!hbt]
    \centering
    \includegraphics[width=0.40\textwidth]{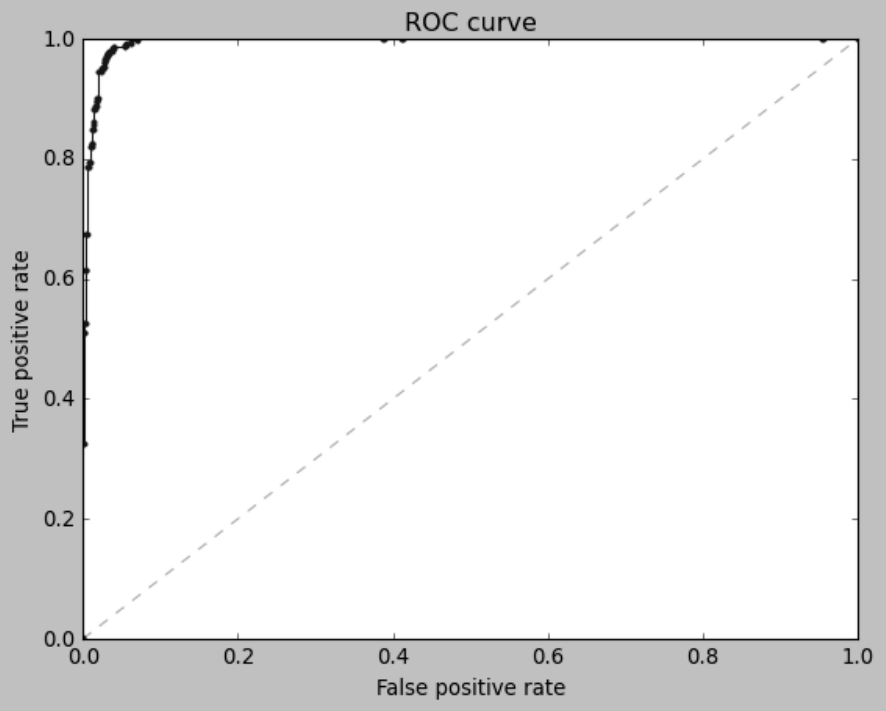}
    \caption{Area Under Curve of the test result}
    \label{fig_auc_cuurve}
\end{figure}

It was noted that the validation or testing accuracy was higher than the training accuracy, and the validation accuracy reached its peak after every 5 epochs, while the validation loss was consistently lower than the training loss. However, at the 4th epoch, the training and testing loss was found to be relatively similar, the model performed exceptionally well on unseen or testing data, which is the major goal for any model. It was also observed that the model did not exhibit overfitting, which is a scenario where the training accuracy is high but the validation or testing accuracy is low. This is further depicted by the confusion matrix in Fig. ~\ref{fig_confusion_matrix} and the Area Under Curve in Fig. ~\ref{fig_auc_cuurve}, both of which provide insightful representations of the model's performance. Additionally, the Table ~\ref{tab:result}  shows the evaluated outcome of the model using the following \eqref{eq_ppv}-\eqref{eq_f1} and Table ~\ref{tab:comparison} shows the performance comparison with some recent related works.

{\footnotesize
\begin{equation}
PPV\ or\ Precision = \frac{TP}{TP + FP} \label{eq_ppv}
\end{equation}

\begin{equation}
TPR\ or\ Recall = \frac{TP}{TP + FN} \label{eq_tpr}
\end{equation}

\begin{equation}
Accuracy = \frac{TP + TN}{TP + TN + FP + FN} \label{eq_acc}
\end{equation}

\begin{equation}
F1-score = 2 \times \frac{PPV \times TPR}{PPV + TPR} \label{eq_f1}
\end{equation}
}

Where, precision measures the exactness of positive predictions, recall assesses the model's ability to identify all positive instances, accuracy reflects overall model performance, and F1 score is the harmonic mean of precision and recall, TP = True Positive, TN = True Negative, FP = False Positive, and FN = False Negative. 

\begin{table}[!hbt]
\centering
\caption{Performance of the model on the test set}
\label{tab:result}
\begin{tabular}{lcc}
\hline
\multicolumn{1}{c}{\textbf{Model}}     & \textbf{Evaluation Metrics} & \textbf{Achieved Scores} \\ \hline
 & Precision   & 0.9149 \\
 & Recall      & 0.9944 \\
\multicolumn{1}{c}{Custom VGG19 model} & F1-Score                    & 0.9530                   \\
 & Specificity & 0.9780 \\
 & Accuracy    & 0.9642 \\ \hline
\end{tabular}
\end{table}

\begin{table}[!hbt]
\centering
\caption{A comparative study with the related works}
\label{tab:comparison}
\resizebox{\columnwidth}{!}{%
\begin{tabular}{cc}
\hline
\textbf{Existing Method} &
  \textbf{Accuracy} \\ \hline
  \begin{tabular}[c]{@{}c@{}}Microscopic brain tumor detection and \\ classification using 3D CNN and feature \\ selection architecture \cite{Rehman}.\end{tabular} &
  98.32\%,  96.97\%, 92.67\% \\ \hline
\begin{tabular}[c]{@{}c@{}}Transfer Learning Based Brain Tumor \\ Detection and Segmentation using Super \\ pixel Technique \cite{Ahuja}.\end{tabular} &
  99.3\% \\ \hline
\begin{tabular}[c]{@{}c@{}}Brain Tumor detection using VGG19 with \\ Fine Tuning \cite{Swati}.\end{tabular} &
  96.13\% \\ \hline
\begin{tabular}[c]{@{}c@{}}Brain tumor classification in MRI image \\ using super pixels, principal component \\ analysis and template-based K-means \\ clustering algorithm \cite{Islam}.\end{tabular} &
  95\% \\ \hline
\begin{tabular}[c]{@{}c@{}}Brain Tumor Analysis Using Deep Learning \\ and VGG-16 Ensemble Learning \\ Approaches \cite{Younis}.\end{tabular} &
  89.5\%, 97.6\%, 91.29\% \\ \hline
Brain tumor detection using VGG-19 \cite{Irmak}  &
  96.32\% \\ \hline
\textbf{\textit{The method of this study}} &
  96.42\% \\ \hline
\end{tabular}%
}
\end{table}

Systematic observation of the previous literature reveals that the methodologies employed for brain tumor classification exhibit substantial differences when compared to our proposed approach. It is noted that a significant number of researchers have relied solely on the VGG-19 architecture without making any modifications, while others have incorporated machine learning algorithms in conjunction with VGG-19. In contrast, our method involves the modification of the VGG-19 architecture through the incorporation of dense layers and dropout layers, leading to a notable improvement in classification accuracy. While there have been several existing methods for brain tumor detection, our work specifically focuses on brain tumor classification and achieved an accuracy of 96.42\%.

\section{Conclusion}
 The overarching objective of this study was to develop an AI-based system that could serve as a decision-support tool for clinicians in detecting brain tumors and enable patients to comprehend the diagnostic reports. To achieve this aim, we employed a CNN architecture based on VGG19, which was modified with additional layers to minimize computational complexity while enhancing performance. Our model was trained on an extensive dataset of MRI images, encompassing both previous data as well as real-time and augmented data, which allowed us to avoid overfitting and surpassing the accuracy of previous models. Moving forward, we plan to expand our dataset and explore various image segmentation methods, with the goal of increasing the scalability and interpretability of the proposed approach.

\end{document}